\documentclass[%
preprint,
aps,
]{revtex4-1}
\usepackage[dvipdfmx]{graphicx,color}
\usepackage{subfigure,
mathrsfs,braket,amsmath,amssymb,dcolumn,bm,comment}
\newlength{\subfigwidth}
\newlength{\subfigcolsep}
\setkeys{Gin}{width=\subfigwidth}
\makeatletter 
\@addtoreset{figure}{section}
\makeatother 
\begin{document}
%\preprint{}
\def\tbr{\textcolor{red}}
\def\tcr{\textcolor{red}}
\def\ov{\overline}
\def\dprime{{\prime \prime}}
\def\nn{\nonumber}
\def\f{\frac}
\def\p{\partial}
\def\H{\mathcal{H}}
\def\beq{\begin{equation}}
\def\eeq{\end{equation}}
\def\bea{\begin{eqnarray}}
\def\eea{\end{eqnarray}}
\def\bsub{\begin{subequations}}
\def\esub{\end{subequations}}
\def\dc{\stackrel{\leftrightarrow}{\partial}}
\def\d{\partial}
\def\sla#1{\rlap/#1}
\def\mH{\mathscr{H}}
\def\tD{\tilde{D}}
\def\Q{{\cal Q}}
\def\mpim{m_{\pi^-}}
\def\mpi0{m_{\pi^0}}
\def\meta{m_\eta}
\def\tauepp{\tau^- \to \nu_\tau \eta \pi^- \pi^0}
\def\Vcurrent{\bar{d}\gamma_\mu u}
\def\Acurrent{\bar{d} \gamma_\mu \gamma_5 u}
\def\VmA{\bar{d}\gamma_\mu(1-\gamma_5)u}
\def\epp{\eta \pi^- \pi^0}
\def\hg{{G}_{2p}}
\def\IP{\mathrm{IP}}
\def\hc{\mathrm{h}.\mathrm{c}.}
%%%%%%    TEXT START    %%%%%%
\def\nn{\nonumber}
\def\beq{\begin{equation}}
\def\eeq{\end{equation}}
\def\bei{\begin{itemize}}
\def\eei{\end{itemize}}
\def\bea{\begin{eqnarray}}
\def\eea{\end{eqnarray}}
\def\aTr{{\rm Tr}}
\def\s{\partial \hspace{-.47em}/}
\def\ad{\overleftrightarrow{\partial}}
\def\para{%
\setlength{\unitlength}{1pt}%
\thinlines %
\begin{picture}(12, 12)%
\put(0,0){/}
\put(2,0){/}
\end{picture}%
}%
%%%%%%%%%%%%%%%%%%
% inserted
%%%%%%%%%%%%%%%%%%
%\newlength{\subfigwidth}
%\newlength{\subfigcolsep}
%\setlength{\subfigcolsep}{2\tabcolsep}
%\setkeys{Gin}{width=\subfigwidth}
%\makeatletter 
%\renewcommand{\thefigure}{% 
%\thesection.\arabic{figure}} 
%\@addtoreset{figure}{section} 
%\makeatother
%%%%%%%%%%%%%%%%%
%%%%%%    TEXT START    %%%%%%
\title{Creation and evolution of particle number
asymmetry in an expanding universe
}
%%%%%%%%%%%%%%%
%%% version  %%%%%
%%%%%%%%%%%%%%%
%Ver.2
\author{Takuya Morozumi}
\email[E-mail: ]{morozumi@hiroshima-u.ac.jp}
\affiliation{Graduate School of Science, Core of Research for Energetic Universe, Hiroshima University, Higashi-Hiroshima, 739-8526, Japan}
\author{Keiko I. Nagao}
\email[E-mail: ]{nagao@sci.niihama-nct.ac.jp}
\affiliation{National Institute of Technology, Niihama College, 792-8580, Japan}
\author{Apriadi Salim Adam}
\email[E-mail: ]{apriadiadam@hiroshima-u.ac.jp}
\affiliation{Graduate School of Science,Core of Research for Energetic Universe, Hiroshima University, Higashi-Hiroshima, 739-8526, Japan}
\author{Hiroyuki Takata}
\email[E-mail: ]{takata@tspu.edu.ru}
\affiliation{Tomsk State Pedagogical University, Tomsk, 634061, Russia}
\date{\today}
%%%%%%%%%%%%%%%
%   Abst %%%%%%%%%
%%%%%%%%%%%%%%%
\begin{abstract}
We  introduce  a  model  which  may generate  particle  number  asymmetry  in
an expanding  Universe. The model includes  CP  violating  and  particle  number violating interactions.  The  model  consists  of  a  
real  scalar  field  and  a  complex  scalar
field.  Starting with an initial condition
specified by a density matrix, we show
how the asymmetry is created through
the  interaction  and  how  it  evolves at later time. 
We compute the asymmetry using non-equilibrium quantum field theory
and as a first test of 
the model, we study how the asymmetry evolves in the flat limit.  
\end{abstract}
%%%%%%%%%%%%%%%
%%  Abst. end  %%%%
%%%%%%%%%%%%%%%
\maketitle
%%%%%%%%%%%%%%%%
%%%  Intro %%%%%%%%
%%%%%%%%%%%%%%%%
\section{Introduction}
The origin of the particle and anti-particle asymmetry of our universe
has not been identified yet. 
We propose a model of a neutral scalar and a complex scalar.
U(1) charge carried by the complex scalar corresponds to the
particle number. CP and U(1) violating interactions are introduced 
and they generate particle and anti-particle asymmetry.
We study time evolution of particle number using two particle irreducible
(2 PI) formalism combined with density matrix formulation of quantum field theory. It enables us to 
study the time evolution of the particle number starting with
an intial state specified with a density matrix.
In the previous work \cite{Hotta:2014ewa}, the time evolution of the
particle number is computed with mass term which violates the particle number
asymmetry. In contrast to the previous work where
initial asymmetry should be non-zero, in this work, we aim to generate
non-zero asymmetry starting with the zero asymmetry at the beginning.
\section{A model with CP and particle number violating interaction}
In this section, we present a Lagrangian for the model.
We denote $N$ for the neutral scalar and $\phi$ as a complex scalar. 
\bea
&& S=\int d^4 x 
({\cal L}_{\rm free}+{\cal L}_{\rm int.}),\\
&& {\cal L}_{\rm free}=\partial_\mu \phi^\ast \partial^\mu \phi
+\frac{B^2}{2} (\phi^2+\phi^{\ast 2})-m_{\phi}^2 |\phi|^2+\frac{1}{2} (\partial_\mu N \partial^\mu N-m_N^2 N^2), \\
&&
{\cal L}_{\rm int.}= A \phi^2 N + A^\ast \phi^{\ast 2} N
+A_0 |\phi|^2 N, 
\eea
where $A$ is a complex number and the corresponding interaction is CP
violating.  $B$ and $A_\phi$ are real numbers.  
The particle number is related to U(1) transformation,
\bea
\phi^{\prime}(x)=e^{i \alpha} \phi(x).
\eea
N\"{o}ether current related to the transformation is,
\bea
j_\mu(x)&=&i \phi^\dagger \overleftrightarrow{\partial}_\mu \phi,
\eea
and the particle number is given as,
\bea
Q(x^0)&=&\int d^3 x j_0(x).
\eea
The U(1) symmetry is explicitly broken by the terms with the coefficients 
$B$ and $A$.
The particle number asymmetry per unit volume is given by $j_0(x)$
and its  expectation value 
is written with a density matrix as follows,
\bea
\langle j_0(x) \rangle={\rm Tr}(j_0(x)\rho(0)).
\eea
The current density $j_0(x)$ is written with  Heisenberg operators and
$\rho(0)$ is an initial density matrix which specifies the initial 
state by means of statistics.
In this work, we use the equilibrium statistical density matrix
as an initial density matrix. Specifically, it is given as,
\bea
\rho(0)=\frac{e^{-\beta H_0}}{{\rm Tr}(e^{-\beta H_0})},
\eea 
where $\beta$ denotes inverse temperature $\frac{1}{T}$
and $H_0$ is a free Hamiltonian which corresponds to the free part of the Lagrangian ${\cal L}_{\rm free}$.  If three dimensional
space is translational invariant, the expectation value of the
current depends only on time.

It is convenient to write all the fileds in terms of real scalar fields defined as,
\bea
\phi(x)=\frac{\phi_1+i \phi_2}{\sqrt{2}} \quad, \phi_3=N. 
\eea
With the definition, the free part of the Lagrangian is rewritten as,
\bea
&& {\cal L}_{\rm free}=\frac{1}{2} 
(\partial_\mu \phi_i \partial^\mu \phi_i)-\frac{m_i^2}{2} \phi_i^2,\nn \\
&& m_1^2=m_{\phi}^2-B^2,  \quad  m_2^2=m_{\phi}^2+B^2, \quad m_3^2=m_N^2. 
\eea
Non-zero $B^2$ leads to the nondegenerate mass
spectrum for $\phi_1$ and $\phi_2$.
The interaction Lagrangian is written with a complete symmetric tensor 
$A_{ijk}, (i,j,k=1,2,3)$
\bea
{\cal L}_{\rm int}&=&
\sum_{ijk=1}^3 \frac{A_{ijk}}{3} \phi_i \phi_j \phi_k. 
\eea
The
non-zero components of $A_{ijk}$ are written with the couplings for cubic interaction, $A$ and $A_\phi$ as shown in Table \ref{tb:1}.
We also summarize the qubic interactions and their property according to U(1) symmetry and CP symmetry. 
\begin{table}
\begin{center}
\caption{The cubic interactions and their property}
\begin{tabular}{|c|c|} \hline
$A_{113}=\frac{A_0}{2}+{\rm Re}.(A)$ & \\ \hline 
$A_{223}=\frac{A_0}{2}-{\rm Re}.(A)$ & \\ \hline
$A_{113}-A_{223}=2 {\rm Re.}(A)$& U(1) violation \\ \hline
$A_{123}=-{\rm Im}.(A)$ &\small{U(1), CP violation} \\ \hline
\end{tabular}
\label{tb:1}
\end{center}
\end{table} 
\section{2 PI effective action and the expectation value for the current}
\begin{figure}[htbp]
\includegraphics[width=4.0cm]{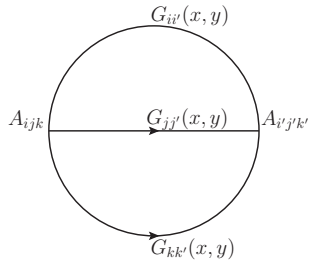}
\caption{Two particle irreducible diagram.}
\label{fig:1}
\end{figure}

The expectation value is written with two parts. 
\bea
\langle j_0(x) \rangle=\lim_{y \rightarrow x}
(\frac{\partial}{\partial x^0}-\frac{\partial}{\partial y^0})
{\rm Re.}[G^{12}_{12}(x,y)]
+ \rm{Re.}\left(\bar{\phi}^\ast_2 \overleftrightarrow{\partial}_0 \bar{\phi}_1 \right),
\eea
where $G^{12}_{12}$ is a Green function and $\overline{\phi}$ is an expectation
value.
$G_{ij}(x,y)$ and $\bar{\phi}_i$ are obtained from 2 PI effective action
$\Gamma[G, \bar{\phi}]$.
\bea  
&& \Gamma[G,\bar{\phi}]=S[\bar{\phi}]+\frac{i}{2}
  \rm{TrLn} G^{-1} +\frac{1}{2}  \int d^{4} x d^{4} y
  \frac{\delta^{2} S}{\delta \bar{\phi}_{i}^{a} ( x ) \delta
  \bar{\phi}_{j}^{b} ( y )} G^{ab}_{ij} (x,y)\nn \\
&+& \frac{i}{3} D_{a b c} A_{i j k} \int \int d^{4} x d^{4} y 
[G^{a a^\prime}_{i i^\prime} ( x,y ) G^{b b^\prime}_{j j^{\prime}} ( x,y )
 G^{c c^\prime}_{k k^{\prime}}(x,y)] D_{a^{\prime} b^\prime c^\prime} A_{i^\prime j^\prime k^\prime}. 
\label{eq:2PI}
\eea
The last term of Eq.(\ref{eq:2PI}) is obtained from two particle irreducible 
diagram shown in Fig. \ref{fig:1}.
\section{expectation value for the current}
While there are  several different contributions to the current up to the first order of the coupling constant $A$, we focus on the contribution which comes 
from Green function. This contribution becomes non-zero even if we start with the vanishing expectation value for the complex scalar as $\bar{\phi_i}(0)=0 (i=1,2)$.
\bea
\langle j(x^0) \rangle^{O(A)}=\lim_{y \rightarrow x}
(\frac{\partial}{\partial x^0}-\frac{\partial}{\partial y^0})
{\rm Re.}[G^{12 O(A)}_{12}(x,y)].
\eea
where $G^{12 O(A)}_{12}$ implies the correction to the first order contribution
with respect to the cubic interaction to the Green function.
We call 
each contribution as, absorption/emission, decay/inverse decay and vacuum.
\bea
\langle j_0(x^0)\rangle^{O(A), \overline{\phi_1}(0)=\overline{\phi_2}(0)=0}
&=&{\langle j_0(x^0)  \rangle}_{absorption}^{emission}+
{\langle j_0(x^0)  \rangle}_{decay}^{inverse \ decay}+ 
{\langle j_0(x^0)  \rangle}_{vacuum}.
\eea 
The contribution corresponds to absorption and emission is, 
\bea
&&{\langle j_0(x^0)  \rangle}_{absorption}^{emission}=
\frac{\overline{\phi_3}(0)A_{123}}{2}
\int \frac{d^3 k}{(2 \pi)^3}\left(\frac{1}{\omega_{1k}}+
\frac{1}{\omega_{2k}} \right) \times \nn \\
&\Biggr{[}& \left(
\frac{\coth \frac{\beta \omega_{2k}}{2}-\coth\frac{\beta \omega_{1k}}{2}
}{2}+\tanh \frac{\beta \omega_{30}}{2}\right) 
\frac{\sin(\omega_{2k}-\omega_{1k})x^0+\sin \omega_{30} x^0}
{\omega_{2k}-\omega_{1k}+\omega_{30}}  \nn \\
&-&
\left(
\frac{\coth \frac{\beta \omega_{1k}}{2}-\coth\frac{\beta \omega_{2k}}{2}
}{2}+\tanh \frac{\beta \omega_{30}}{2}\right)
\frac{\sin(\omega_{1k}-\omega_{2k})x^0+\sin \omega_{30} x^0}
{\omega_{1k}-\omega_{2k}+\omega_{30}}
\Biggl{]}.
\eea
where $\omega_{ik}=\sqrt{k^2+m_i^2} \quad (i=1,2,3)$.
The decay and inverse decay contribution is given as,
\bea
&&{\langle j_0(x^0)\rangle}_{decay}^{inverse \ decay}=
\frac{\overline{\phi_3}(0)A_{123}}{2}
\int \frac{d^3 k}{(2 \pi)^3}
\left(\frac{1}{\omega_{2k}}-\frac{1}{\omega_{1k}}
\right) \times \nn \\
&& \left(
\frac{\coth \frac{\beta \omega_{2k}}{2}+\coth \frac{\beta \omega_{1k}}{2}-2 \tanh \frac{\beta \omega_{30}}{2}}{2}\right)
\frac{\sin \omega_{30} x^0-
\sin(\omega_{1k}+\omega_{2k})x^0}{\omega_{30}-\omega_{2k}-\omega_{1k}}.
%&& \omega_{2k}+\omega_{1k}=\omega_{30} > m_1 + m_2. \nn  
\eea
The vacuum contribution is,
\bea
&&{\langle j_0(x^0)  \rangle}_{vacuum}
= \frac{\overline{\phi_3}(0)A_{123}}{2}
\int \frac{d^3 k}{(2 \pi)^3}
\left(\frac{1}{\omega_{2k}}-\frac{1}{\omega_{1k}}
\right) \times \nn \\
&& \left(
\frac{\coth \frac{\beta \omega_{2k}}{2}+
\coth \frac{\beta \omega_{1k}}{2}+
2 \tanh \frac{\beta \omega_{30}}{2}}{2}\right)
\frac{\sin \omega_{30} x^0+
\sin(\omega_{1k}+\omega_{2k})x^0}{\omega_{30}+\omega_{2k}+\omega_{1k}}.
\eea
\section{Conclusion}
We propose a model of scalars which may generate the particle number asymmetry.
In the interacting model, 2 PI effective action  
$\Gamma[G, \bar{\phi}]$ and 
Schwinger Dyson equation for Green functions $G$ and expectation value 
$\bar{\phi}$ are obtained. They 
are  iteratively solved by treating  interaction
$A_{ijk}$ is small.
The current for the particle and 
anti-particle asymmetry is given up to the first order of $A$.
The contribution is classified to five important processes.
As a future extension of the work, we will carry out the numerical calculation
of the asymmetry. 


\begin{thebibliography}{99}
%\cite{Hotta:2014ewa}
\bibitem{Hotta:2014ewa} 
  R.~Hotta, T.~Morozumi and H.~Takata,
  %``Time variation of particle and antiparticle asymmetry in an expanding universe,''
  Phys.\ Rev.\ D {\bf 90}, no. 1, 016008 (2014)
% doi:10.1103/PhysRevD.90.016008
  [arXiv:1403.0733 [hep-ph]].
  %%CITATION = doi:10.1103/PhysRevD.90.016008;%%

\end{thebibliography}
\end{document}